\documentclass[aps,twocolumn,amsmath,amssymb,longbibliography,prb,10pt]{revtex4-2}
\usepackage[colorlinks=true,allcolors=blue,hypertexnames=false]{hyperref}
\usepackage{amsmath}
\usepackage{dsfont}
\usepackage{textcomp}
\usepackage{tabularx}
\usepackage{graphicx}
\usepackage{soul}
\usepackage{feynmp-auto}
\usepackage{footmisc}
\usepackage[export]{adjustbox}
\usepackage{wrapfig}
\usepackage{bbm}

\makeatletter
\def\@bibdataout@aps{%
 \immediate\write\@bibdataout{%
  @CONTROL{%
   apsrev42Control%
   \longbibliography@sw{%
    ,author="48",editor="1",pages="0",title="0",year="1"%
   }{%
    ,author="48",editor="1",pages="0",title="",year="1"%
   }%
  }%
 }%
 \if@filesw
  \immediate\write\@auxout{\string\citation{apsrev42Control}}%
 \fi
}%
\makeatother

\newcommand{\beg}{\begin{equation}}
\newcommand{\en}{\end{equation}}

\newcommand{\bk}{\mathbf k}

\newcommand{\bn}{\mathbf n}

\newcommand \bel  {\begin{align}}
\newcommand \enl  {\end{align}}

\newcommand{\up}{\uparrow}
\newcommand{\dn}{\downarrow}
\newcommand{\dg}{^\dagger}

\newcommand{\ttau}[1]{\hat\tau_{#1}}

\newcommand{\hs}{\hat\sigma}

\def\XXint#1#2#3{{\setbox0=\hbox{$#1{#2#3}{\int}$}
     \vcenter{\hbox{$#2#3$}}\kern-.5\wd0}}

\begin{document}

\title{Spin-polarized tunneling from a Kondo insulator: equilibrium surface magnetism versus magneto-electric response in SmB$_6$}

\author{Maxim Dzero}
\affiliation{Department of Physics, Kent State University, Kent, Ohio 44242, USA}

\author{Brian M. Andersen}
\affiliation{Niels Bohr Institute, University of Copenhagen, DK-2100 Copenhagen, Denmark}

\date{\today}

\begin{abstract} 
We theoretically analyze the spin-polarized tunneling between SmB$_6$ tip and the bicollinear antiferromagnet Fe$_{1+x}$Te. The emergence of magnetic contrast in tunneling requires time-reversal symmetry breaking on the surface of the tip. The magnetic contrast part of the tunneling conductance then has two contributions: the first one originates from the magnetic moment due to the equilibrium long-range order on the surface, while the other is governed by the magnetic moment induced by the electric field of the junction through the surface magneto-electric effect. We show that the ratio of the equilibrium moment to the response extracted near zero bias equals, up to the sign, twice the bias at which the contrast changes sign, independently of the relative strength of the two mechanisms, so that this ratio itself does not measure the strength of the magneto-electric effect. We find that within the experimental energy window $|eV|\lesssim10$~meV the {bias-dependent} field-induced contribution is at most a few percent of the equilibrium one, and the two become comparable only at biases of order $100$~meV or higher.  The ratio of the field-induced to the equilibrium contribution is inversely proportional to the distance between the tip and the substrate.
\end{abstract}

\maketitle
\section{Introduction}
Samarium hexaboride has been known to host a broad variety of remarkable quantum effects which range from the resistivity saturation \cite{MenthBuehlerGeballe1969,Riseborough2000,Nakajima2016} and current rectification \cite{zhang2025diode} at low temperatures to de Haas - van Alphen oscillations in an apparently insulating state \cite{Li2014,Tan2015,KnolleCooper2015,Erten2016,Zhang2016,Sodemann2018,Shen2018,Hartstein2018,Thomas2019,Li2020,Julian2023} suggesting that a qualitatively new mechanism for quantum oscillations in magnetization must be at play. It is arguably the most studied candidate for a topological Kondo insulator with fully insulating bulk \cite{Dzero2010,Takimoto2011,Alexandrov2013,Dzero2016}. In particular, its low-temperature surface conduction \cite{Wolgast2013,Kim2013,zhang2025diode} and the in-gap surface states observed in photoemission \cite{Xu2013,Jiang2013,Neupane2013}, together with their spin texture \cite{Xu2014}, have been interpreted as evidence for topologically protected Dirac surface states. 

More recently, the surface states in SmB$_6$ have been probed by spin-polarized scanning tunneling microscopy with SmB$_6$ nanowire tips \cite{Aishwarya2022}. Specifically, this tunneling experiment showed quite surprisingly that with the bi-collinear antiferromagnetic order of Fe$_{1+x}$Te used as a substrate, the resulting magnetic contrast reverses when the polarity of the bias is reversed, and it disappears above approximately $10$~K: a temperature at which Cr tips still image the antiferromagnetic order. These experimental observations were initially interpreted as a consequence of the spin-momentum locking of the Dirac surface states \cite{Aishwarya2022}.

\begin{figure*}
\includegraphics[width=0.95\linewidth]{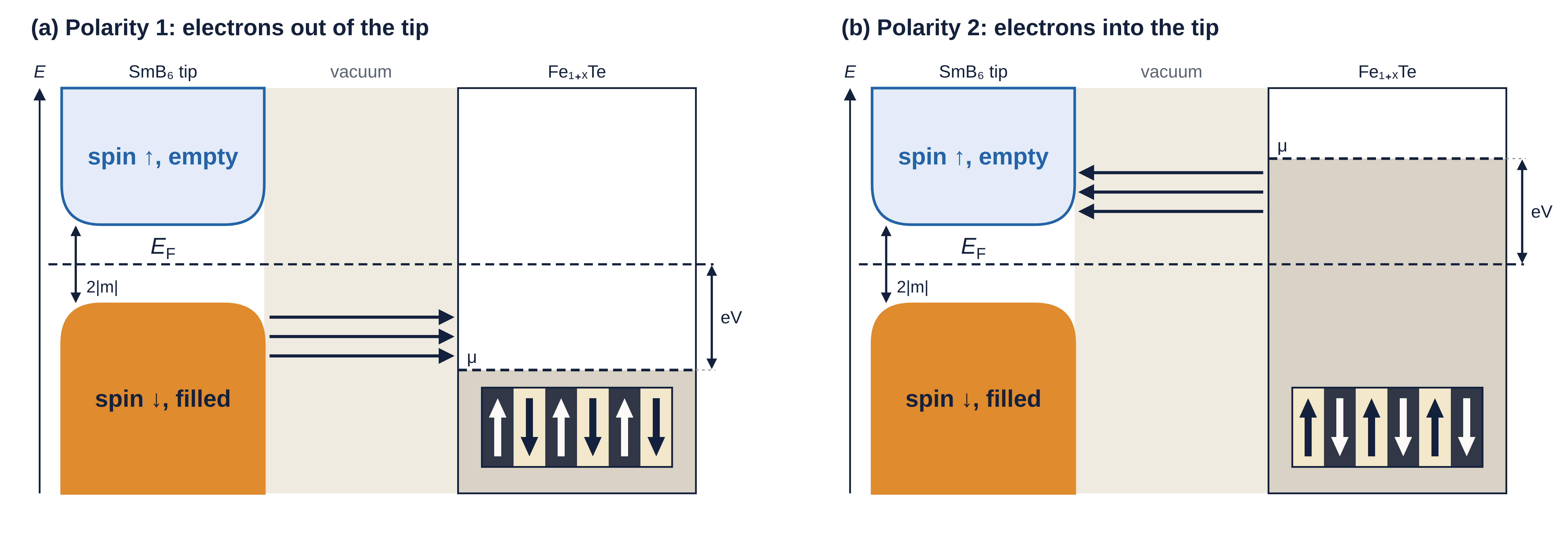}
\caption{Schematic representation for the microscopic origin of the bias reversal of the magnetic contrast for a spin-polarized apex in equilibrium \cite{Brian2026}. The time-reversal-breaking surface order splits the apex states at the Fermi level $E_F$ into a filled spin-$\downarrow$ band below $E_F$ and an empty spin-$\uparrow$ band above it, separated by $2|m|$ (left part of each panel). The Fe$_{1+x}$Te substrate is an antiferromagnet whose chemical potential $\mu$ is shifted by the bias $eV$ relative to $E_F$. (a) For one polarity electrons leave the tip: the current is carried by the filled spin-$\downarrow$ states of the apex into empty substrate states, so the contrast is set by the spin-
$\downarrow$ spectral weight at $E_F-eV$. (b) For the opposite polarity electrons enter the tip from the filled substrate states into the empty spin-$\uparrow$ states of the apex, and the contrast is set by the spin-$\uparrow$ weight at $E_F+eV$. The two polarities therefore sample opposite spin projections of the same equilibrium spectrum $\rho_{\rm spin}(\omega)$, and the magnetic contrast, which is proportional to the projection of the substrate moments on the apex spin, changes sign between them. The same reversal takes place within one polarity when $eV$ crosses a sign change of $\rho_{\rm spin}(\omega)$, at $V^*$.}
\label{Fig-polarity}
\end{figure*}

It has since been shown that no time-reversal symmetric tip can produce such magnetic contrast, let alone its reversal with the polarity of the bias, first for a point-like apex coupled to helical surface states \cite{Banerjee2025} and then most recently for a tip of an arbitrary shape as well as an arbitrary single-particle spectrum of the interacting or non-interacting system alike \cite{Brian2026}. 
Thus these theoretical ideas combined with experimental observations imply that time-reversal symmetry must be broken at the surface of the SmB$_6$ tip below approximately $10$~K, which is well above the Curie temperature of about $0.6$~K of the surface ferromagnetism reported earlier in transport measurements \cite{Nakajima2016}. What makes these observations even more surprising is that SmB$_6$ is a mixed-valence compound \cite{Cohen1970,Chazalviel1976,Mizumaki2009} which does not order magnetically in the bulk at ambient pressure \cite{Gheidi2019} with the long-range magnetic order appearing only above about $6$~GPa, where the charge gap closes and the system becomes metallic in the bulk \cite{Barla2005,Derr2008}.

Two interpretations of the bias reversal have been put forward recently. Banerjee \emph{et al.} \cite{Banerjee2025}, who were the first to link this magnetic contrast to time-reversal symmetry breaking, attributed the bias reversal to a magnetic moment of the tip induced by the electric field of the junction through the surface magneto-electric effect, and extracted the corresponding response from the data assuming a bias-independent equilibrium moment. In Ref.~\cite{Brian2026} we have shown that an equilibrium time-reversal-breaking surface order gaps out the surface Dirac cones and makes the spin-resolved spectral function of the apex change sign close to zero bias, so that the bias reversal occurs without any field-induced moment. We provide a schematic illustration of this idea in Fig.~\ref{Fig-polarity}.

The time-reversal symmetry breaking on the surface of SmB$_6$ which we described with the fully microscopic model \cite{Brian2026} is odd under both time reversal and inversion. It therefore also allows for a linear magneto-electric response \cite{Wilczek1987,Qi2008,Essin2009,Mogi2017,Sekine2021,Li2023}. Thus the equilibrium and the field-induced moments of the apex are both present in the junction, and so the natural question arises as to which one of the two provides the dominant contribution to the magnetic contrast. In this paper we address this question within the microscopic theory which we developed in Ref.~\cite{Brian2026}. Our theory allows us to compute the magneto-electric and the equilibrium contributions to the spin-resolved tunneling conductance within the same methodological framework and then compare them. Specifically, we compute the change of the spin-resolved spectral function of the apex to linear order in the bias. This is justified since the {bias-dependent part of the} electric field couples to the apex with energies of the order of $10~\mu$eV per millivolt of bias, which is small compared with the width of the spectral features of the apex. {The static field set by the contact potential of the junction is larger by one to two orders of magnitude. Crucially, it does not reverse with the bias polarity and therefore it can be absorbed into the equilibrium state about which we expand (see Sec.~\ref{sec:discussion}).} The linear response itself, however, need not be small, since it includes the field-induced change of the order parameter, which grows as the temperature approaches $T_c$. We find that within the experimental window $|eV|\lesssim10$~meV the magneto-electric contribution is at most a few percent of the equilibrium one, and the two become comparable only at biases of order $100$~meV. It thus follows that both the energy dependence of the magnetic contrast and its reversal with bias are determined, to within a few percent, by the equilibrium spin-resolved spectrum of the apex.

Our paper is organized as follows. In the next Section we provide the estimates for the relative contributions to the spin-resolved differential conductance from the equilibrium spectrum and magneto-electric effect. Section III formulates the theoretical framework which allows us to compute the net tunneling conductance and summarizes the main results of this work. Sections IV and V are devoted to the discussion and conclusions correspondingly. A few technical aspects of our work are given in Appendix A. 

\section{Qualitative discussion}
To the lowest order in tunneling, the part of the current which changes sign when the tip is moved by half of the magnetic lattice period of the Fe$_{1+x}$Te substrate is \cite{Brian2026}:
\beg\label{Eq1}
\begin{aligned}
\delta I^{\rm odd}(V)\propto &\int d\omega[f(\omega+eV)-f(\omega)]\\&\times
\rho_{\rm spin}(\omega+eV)(\hat{\mathbf n}\cdot\hat{\mathbf z}) \rho_1(\omega).
\end{aligned}
\en
Here the scalar product appears as a result of taking the trace over the corresponding Pauli matrices and it involves unit vectors $\bn$ which define the direction of the magnetic moments in the substrate and $\hat{\mathbf z}$ is a normal to the plane of the apex. Furthermore,  
$\rho_1(\omega)$ is the spin-polarization function of the electrons in the substrate and $\rho_{\rm spin}=\rho_{\up}(\omega)-\rho_{\dn}(\omega)$ is the spin-resolved spectral function projected on the surface normal of the apex. In what follows without loss of generality we will assume that  $\rho_1(\omega)$ is a smooth function of energy on the scale of the apex structure.
Then \eqref{Eq1} can be re-written as $\delta I^{\rm odd}(V)=-{\cal C}{J}_{\rm odd}(V)$ where
\beg\label{tIodd}
{J}_{\rm odd}(V)=\int d\omega[f(\omega-eV)-f(\omega)]
\rho_{\rm spin}(\omega,V),
\en
$f(\omega)$ is the Fermi-Dirac distribution function and ${\cal C}\propto\rho_1(\hat{\mathbf n}\cdot\hat{\mathbf z})$ is a constant which will be of no importance for our subsequent discussion. 

In Ref. \cite{Brian2026} the dependence of  $\rho_{\rm spin}(\omega,V)$ on the bias voltage has been completely ignored.  However in a tunneling junction $\rho_{\rm spin}$ generally does depend on the applied bias through its dependence on electric field. Specifically, the bias drop across the gap between the tip and the substrate leads to the appearance of an electric field 
\beg\label{Efield}
E=\frac{V+V_c}{d}, 
\en
where $d$ is the distance between the electrodes and $V_c$ is an additional bias due to the difference between the corresponding work functions of the electrodes. The latter is clearly there even when $V=0$. However mainly for convenience we will set it to zero in what follows.  Thus, to linear order in the electric field we then write
\begin{equation}\label{eq:split}
\rho_{\rm spin}(\omega,V)=\rho_{\rm s}(\omega)+V\,X(\omega),
\end{equation}
where $\rho_{\rm s}(\omega)$ is the equilibrium spin-resolved spectral function which has been evaluated in Ref.  \cite{Brian2026} and $X(\omega)\propto 1/d$ collects every channel by which the electric field changes the single-particle spectrum in the apex. We consider three such channels: the on-site $d$--$f$ electric dipole, electrostatic gating, and the field-induced change of the order parameter:
\beg\label{SplitB}
X(\omega)=X_{d}(\omega)+X_{\rm g}(\omega)+X_\Delta(\omega).
\en
In the spectral sense all of them are magneto-electric in origin, since each gives a spin spectral weight linear in $E$ and odd under reversal of the magnetic domain \cite{Wilczek1987,Qi2008,Essin2009,Mogi2017,Sekine2021,Li2023,Banerjee2025}. {Lastly, we would like to emphasize that setting $V_c=0$ in Eq.~\eqref{eq:split} does not mean that the contact field is neglected: $\rho_{\rm s}(\omega)$ is to be understood as the spectral function of the apex in the static field $V_c/d$, and $X(\omega)$ describes its change under the additional field of the bias. Since $|V|\lesssim50$~mV $\ll|V_c|$ in the experiments \cite{Aishwarya2022,Banerjee2025}, the latter is a small perturbation on top of the former.}

Two remarks on Eq.~\eqref{eq:split} are in order. First, as we have shown in Ref. \cite{Brian2026}, the equilibrium part $\rho_{\rm s}(\omega)$ is itself odd under time reversal: it is the spin-resolved spectral function of an apex whose surface order breaks time-reversal symmetry. Moreover $\rho_{\rm s}(\omega)$ is an energy-dependent function with sign changes of its own, and the tunneling conductance samples it at $\omega=eV$. The reversal of the magnetic contrast with bias is therefore already contained in $\rho_{\rm s}(\omega)$ with no electric field required and this alone may give rise to the emergence of magnetic contrast in tunneling as we illustrate in Fig.~\ref{Fig-polarity}. Second, the analysis of Ref.~\cite{Banerjee2025} is based on representing the tip moment as $m_t(\omega,V)=m_0+V X(\omega)$ with $m_0$ a {bias-independent number}, so that all of the energy dependence of the magnetic contrast has been attributed to the field-induced part $X(\omega)$. For a magnetically ordered apex, however, the equilibrium part $\rho_{\rm s}(\omega)$ is a strongly energy-dependent function \cite{Brian2026}, and an inversion that subtracts only a constant returns the energy dependence of $\rho_{\rm s}(\omega)$ as part of the extracted magneto-electric contribution $X(\omega)$. We make this statement precise in Sec.~\ref{sec:gspin} below, where we show that the extracted function equals $X(\omega)$ plus a term generated by $\rho_{\rm s}(\omega)$ alone, and that the ratio of the equilibrium moment to the extracted response near zero bias is fixed by the position of the sign change, independently of the relative size of the two contributions.

The first two channels in Eq.~\eqref{SplitB} are responses of the single-particle spectrum at fixed order parameter. Their magnitude is set by the field at the apex, by the corresponding length scale and by the derivative of the spectrum. The on-site $d-f$ dipole (see Appendix \ref{AppendixG8} for details) mixes the even-parity $5d$ and the odd-parity $4f$ states of the apex ion with the coupling $\lambda=eE_{\rm apex}{\mathsf z}_{\rm df}$, where $E_{\rm apex}$ the electric field the apex sees, ${\mathsf z}_{\rm df}=\langle d|\hat z|f\rangle$ is the intra-atomic dipole length, of order $0.1$~\AA. The electrostatic potential of the apex layer is $\phi_0=E_{\rm apex}\ell_{\rm s}$, where $\ell_{\rm s}\sim$ $10^{-2}$~\AA\ is the screening length of the metallic surface layer. Writing $E_{\rm apex}=(V+V_c)/(\kappa d)$ with a screening factor $\kappa>1$ that accounts for the partial penetration of the field into the apex, from Eq. \eqref{eq:split} we have
\beg\label{Bdirect}
X_{d}(\omega)+X_{\rm g}(\omega)=\frac{e \ell_{\rm s}}{\kappa d}\left[\frac{{\mathsf z}_{\rm df}}{\ell_{\rm s}}\,\frac{\partial\rho_{\rm s}}{\partial\lambda}+\,\frac{\partial\rho_{\rm s}}{\partial (e\phi_0)}\right]_{\Delta_j,h_j},
\en
where the derivatives are taken at fixed order-parameter profile $(\Delta_j,h_j)$ through the layers in the tip labeled by $j$ \cite{Brian2026} and $\phi_0$ is the potential of the apex layer. The derivatives are of order $1$~eV$^{-1}$ per meV, the same order as $\rho_{\rm s}/\Delta$, while the energies $eE_{\rm apex}{\mathsf z}_{\rm df}$ and $eE_{\rm apex}\ell_{\rm s}$ are of order $10~\mu$eV per millivolt of bias. For the unscreened field at $d=7$~\AA\ the two terms of Eq.~\eqref{Bdirect} at $\omega=1$~meV are $-0.008$ and $-0.009$~eV$^{-1}$ per mV against $\rho_{\rm s}(1~\text{meV})=1.66$~eV$^{-1}$: about one percent of the equilibrium spectrum per millivolt, so that within the window $|eV|\lesssim10$~meV these channels change $\rho_{\rm spin}$ by at most a few percent and would become comparable to $\rho_{\rm s}$ only for $|eV|$ of order $100$~meV. Screening, $\kappa>1$, can only reduce these estimates. For this reason the corresponding contribution to the differential tunneling conductance from these two channels will be ignored in what follows. 

The third channel contribution $X_\Delta(\omega)$ is quite different from the first two. It originates from the process in which  the electric field in the junction acts on the many-body ordered state described by the order parameter $\psi=(\Delta_{PT},h)$ through the coupling to the gating potential $e\phi_0$ and to the  $d$--$f$ dipole  $\lambda$. This response is governed by the amplitude susceptibility of the surface time-reversal symmetry breaking order, whose inverse vanishes at $T_c$. Therefore $X_\Delta(\omega)/\rho_s(\omega)$ being proportional to the susceptibility carries a factor $(T_c-T)^{-1}$ in mean field and is not bounded by the smallness of the couplings, so that near $T_c$ it must give a dominant contribution to $X(\omega)$. Well below $T_c$ the enhancement is approximately equal to $\alpha T_c/(T_c-T)$ with $\alpha\sim 10$. Since the same couplings drive all three channels, $X_\Delta(\omega)$ is expected to be smaller than the first two away from $T_c$ and comparable to them close to $T_c$, with a relative sign that no symmetry fixes, since all three are odd under reversal of the domain. Importantly, both the sign and the magnitude of $X_\Delta(\omega)$ can only be found from the microscopic calculation, which we present in what follows.

\section{Magnetic contrast part of the differential conductance}\label{sec:gspin}
\subsection{Equilibrium and field-induced contributions at fixed distance between the tip and the substrate}
Let us introduce the magnetic contrast part of the differential conductance $g_{\rm spin}(V)\equiv dJ_{\rm odd}/dV$. Given this definition and using  Eq.~\eqref{tIodd} along with the spectral function~\eqref{eq:split} yields
\begin{align}\label{gspin}
g_{\rm spin}(V)&=g_{\rm s}(V)+{\cal G}(V)+eV\overline{X}(eV),
\end{align}
where $g_s(V)=e\rho_{\rm s}(eV)$ is the spin-resolved differential tunneling conductance originating from the equilibrium spectrum. The function ${\cal G}(V)$ is defined according to
\beg\label{jV}
\begin{aligned}
&{\cal G}(V)\equiv\int d\omega\,[f(\omega-eV)-f(\omega)]\,X(\omega), 
\end{aligned}
\en
and $\overline{X}(eV)$ is found by the frequency convolution of the function $X(\omega)$ with the temperature dependent function $(-\partial f/\partial\omega)_{\omega-eV}$. Thus the field-induced part contributes both through its value at the bias energy and through the integral ${\cal G}(V)$. 

It is instructive to consider the case of  $T=0$. We have
\beg\label{gT0}
{g_{\rm spin}(V)}=g_{\rm s}(V)+eV\cdot X(eV)+\int\limits_0^{eV}X(\omega)\,d\omega,
\en
that for $X=0$ reproduces our result of Ref.~\cite{Brian2026}. Expanding Eq.~\eqref{gT0} at small bias, the sign change of the contrast is at
\beg\label{Vstar}
V^*\simeq-\frac{g_{\rm s}(0)}{g_{\rm s}'(0)+2eX(0)},
\en
so that the slope of the equilibrium contribution at zero bias and the electric field response enter only through the combination $g_{\rm s}'(0)+2eX(0)$. This means that if we replace $X\to X+\delta X$ and 
also 
\begin{align}
g_{\rm s}(\omega)\to g_{\rm s}(\omega)-\omega\delta{X}(\omega)-\int\limits_0^\omega\delta{X}(\omega')\,d\omega'\label{degeneracy}
\end{align}
then this leaves $g_{\rm spin}(V)$ unchanged at every bias at $T=0$ for any function $\delta{X}(\omega)$. At finite temperature any correction 
$\delta g_{\rm s}(\omega)$ to $g_{\rm s}(\omega)$, Eq.\eqref{degeneracy}, acquires a temperature-dependent correction and the degeneracy persists. 
We therefore conclude that a contrast spectrum taken at a fixed distance between the tip and the substrate cannot separate the equilibrium contribution from the field-induced part of the contribution of the apex magnetism.

Equation~\eqref{degeneracy} has a direct consequence for the inversion of the experimental data. To see this we insert Eq.~\eqref{eq:split} into Eq.~\eqref{tIodd} at $T=0$ and divide both sides by $eV$. If the equilibrium part were a constant, $\rho_{\rm s}(\omega)=m_0$, as it was assumed in Ref.~\cite{Banerjee2025}, we would find $X(eV)=d[J_{\rm odd}(V)/eV]/dV$. For an energy-dependent $\rho_{\rm s}(\omega)$ the same operation gives
\begin{equation}\label{Xext}
X_{\rm act}(eV)=X(eV)+\frac{\rho_{\rm s}(eV)}{V}-\frac{1}{eV^2}\int\limits_0^{eV}\rho_{\rm s}(\omega)\,d\omega,
\end{equation}
so that only a constant equilibrium spectrum is removed, while its energy dependence reappears as an apparent field-induced response. Near zero bias the last two terms in Eq.~\eqref{Xext} tend to $(1/2)(d\rho_{\rm s}/dV)_{V=0}$, so that for $V_c=0$ the denominator of Eq.~\eqref{Vstar} equals $2eX_{\rm act}(0)$ and its numerator equals $em_0$, where $m_0=\rho_{\rm s}(0)$ is the zero-bias intercept. Thus Eq.~\eqref{Vstar} can be re-written in much simpler form as $V^*=-m_0/[2X_{\rm act}(0)]$.

\subsection{Model}\label{Model}
In order to setup the calculation of the magneto-electric contribution to the differential tunneling conductance, we will mainly follow the avenue of Ref. \cite{Brian2026}. In particular, we consider the apex with the (001) termination and use  the eight-orbital model of SmB$_6$ \cite{Roy2014} defined for $N$ explicit layers $j=0,\dots,N-1$ on the semi-infinite bulk. The corresponding mean-field Hamiltonian reads ~\cite{Brian2026}:
\beg\label{Hmf}
\hat{H}_{\rm mf}=\hat{H}_{\rm slab}+\sum_j\big[\Delta_j\,\hat{\cal O}_\Delta(j)+h_j\,\hat{\cal O}_M(j)\big],
\en
which acts in the basis $\Phi=[\Phi_d,\Phi_f]^T$ defined by
\begin{equation}\label{basis}
\Phi_l=\left[l_{1\uparrow}, l_{1\downarrow}, l_{2\uparrow}, l_{2\downarrow} \right]^T,
\end{equation}  
with $l=d,f$, and $f_{j\up,\dn}$ refer to the components of the $j$-th Kramers doublet. 
In \eqref{Hmf} we introduced the operators 
\beg\label{ODLTOM}
\begin{aligned}
\hat{\cal O}_\Delta&=\ttau{2}\otimes\hs_0\otimes\hat{\gamma}_0,\\
\hat{\cal O}_M&=\tfrac12\left(\ttau{0}-\ttau{3}\right)\otimes\hs_z\otimes\hat{\gamma}_0,
\end{aligned}
\en
where $\hat{\gamma}_0$ is a unit matrix and $\hat{\gamma}_3$ is the third Pauli matrix which acts in the orbital subspace. Lastly, $\hat{H}_{\rm slab}$ acts on the bulk states \cite{Brian2026}. 

Since we are interested in studying the effects associated with the electric field at the junction we need to add the following junction Hamiltonian to our model Hamiltonian \eqref{Hmf}
\beg\label{Hext}
\begin{aligned}
\hat{H}_{\rm ext}=&\sum\limits_j\left[e\phi_j\,\hat{\cal O}_\phi+\lambda_j\,\hat{\cal O}_\lambda\right],
\end{aligned}
\en
where we introduced the operators
\beg\label{Os}
\begin{aligned}
\hat{\cal O}_\phi&=\ttau{0}\otimes\hs_0\otimes\hat{\gamma}_0,\\
\hat{\cal O}_\lambda&=\ttau{2}\otimes\hs_z\otimes\hat{\cal P}_2,
\end{aligned}
\en
and
a projection operator onto the $\Gamma_8^{(2)}$ Kramers doublet $\hat{\cal P}_2=(\hat{\gamma}_0-\hat{\gamma}_3)/2$. In what follows we will consider the case when the external field acts on the apex only, so that $\phi_j=E_{\rm apex}\ell_{\rm s}\delta_{j,0}$ and $\lambda_j=eE_{\rm apex}{\mathsf z}_{\rm df}\delta_{j,0}\equiv\lambda \delta_{j,0}$, Eq. \eqref{Bdirect}. For details on the origin of the second term in \eqref{Hext}, we refer the reader to Appendix \ref{AppendixG8}. 

In what follows we will treat \eqref{Hext} as a perturbation and compute the corrections to the order parameters to linear order in powers of $e\phi_0$ and $\lambda$ using the self-consistency equations $\Delta_j=-U_{\rm df}\langle\hat{\cal O}_\Delta(j)\rangle$ and $h_j=-I_{\rm ff}\langle\hat{\cal O}_M(j)\rangle$ with
\beg\label{Self-consistency}
\begin{aligned} 
\langle\hat{\cal O}_\Delta\rangle&=T\sum\limits_{i\omega_n}\int\frac{d^2\bk_\parallel}{(2\pi)^2}{\rm Tr}\left[\hat{\cal O}_\Delta\hat{G}_{jj}(\bk_\parallel,i\omega_n)\right], \\ 
\langle\hat{\cal O}_M\rangle&=T\sum\limits_{i\omega_n}\int\frac{d^2\bk_\parallel}{(2\pi)^2}{\rm Tr}\left[\hat{\cal O}_M\hat{G}_{jj}(\bk_\parallel,i\omega_n)\right].
\end{aligned}
\en
Here we suppressed the dependence of the operators $\hat{\cal O}_{\Delta,M}$ on the layer index for brevity, $\omega_n=\pi T(2n+1)$ is a fermionic Matsubara frequency. In \eqref{Self-consistency} we also introduced the single-particle propagator in the Matsubara representation \cite{AGD}
\beg\label{Gprop}
\hat{G}_{ij}(1,2)=-\frac{\left\langle\hat{T}_\tau\left\{\hat{\Psi}_{i}(1)\hat{\Psi}_{j}\dg(2)\hat{\cal S}_{\rm ext}(\beta)\right\}\right\rangle_{\rm mf}}{\left\langle\hat{\cal S}_{\rm ext}(\beta)\right\rangle_{\rm mf}},
\en
where we used the convention $(1)=(\bk_{\parallel},\tau)$, $(2)=(\bk_{\parallel},\tau')$, $\hat{\Psi}_j(\bk_{\parallel},\tau)$ is a fermionic field on layer $j$ which corresponds to the states given by \eqref{basis} and
\beg\label{Sext}
\hat{\cal S}_{\rm ext}(\beta)=\hat{T}_\tau e^{-\int\limits_0^\beta \hat{H}_{\rm ext}(\tau)d\tau}.
\en
In Eq. \eqref{Gprop} the fermionic operators are considered in the interaction representation and the thermal average is computed with the mean-field Hamiltonian $\hat{H}_{\rm mf}$. 

We now turn our discussion to the computation of the function $X_\Delta(\omega)$, Eq.\eqref{SplitB}.  In full analogy with \eqref{Bdirect}, for the function $X_\Delta(\omega)$ we write 
\beg\label{BDelta}
\begin{aligned}
X_\Delta(\omega)&=\sum_{j=0}^{N-1}\left[\frac{\partial\rho_{\rm s}(\omega)}{\partial\Delta_j}\frac{d\Delta_j}{dV}
+\frac{\partial\rho_{\rm s}(\omega)}{\partial h_j}\frac{dh_j}{dV}\right].
\end{aligned}
\en
In order to keep the subsequent formulas in a compact form, we use the vectors ${\vec \psi}_j=(\Delta_j,h_j)$ and ${\vec \eta}=(U_{\rm df},I_{\rm ff})$. The electric field induced correction to the order parameter components can be compactly written as follows
\beg\label{dpsidV}
\frac{d{\vec \psi_j}}{dV}=\frac{e\ell_{\rm s}}{\kappa d}\left[\frac{{\mathsf z}_{\rm df}}{\ell_{\rm s}}\,\frac{\partial{\vec \psi}_j}{\partial\lambda}+\frac{\partial{\vec \psi}_j}{\partial (e\phi_0)}\right].
\en

The derivatives entering here can be computed by perturbation theory. Given the form of the expressions above, we will only need the correction to the order parameter up to the first order in $e\phi_0$ and $\lambda$, so we write 
\beg\label{Expandpsij}
{\vec \psi}_j\approx{\vec \psi}_j^{(0)}+{\vec \psi}_j^{(1)}\, ,
\en
where ${\vec \psi}_j^{(0)}$ is the solution of the self-consistency equations \eqref{Self-consistency} in the absence of external field. Given the form of the model Hamiltonian \eqref{Hmf}, we also need to include the corresponding linear correction to the order parameter into the definition of $\hat{H}_{\rm ext}$
\beg\label{HextLin}
\hat{H}_{\rm ext}\to \hat{H}_{\rm ext}+\sum_j\big[\Delta_j^{(1)}\,\hat{\cal O}_\Delta(j)+h_j^{(1)}\,\hat{\cal O}_M(j)\big].
\en
Next we expand \eqref{Sext} with \eqref{HextLin} up to the first order in $e\phi_0$, $\lambda$ and ${\vec \psi}_j^{(1)}$.  Inserting the resulting expressions into Eqs. (\ref{Self-consistency},\ref{Gprop}), we derive the following system of linear equations
\beg\label{LinCorr}
\begin{aligned}
\Delta_j^{(1)}&=U_{\rm df}\left\{(e\phi_0)\,\chi_{\Delta\phi}(j,0)+\lambda\,\chi_{\Delta\lambda}(j,0)\right.\\&\left.+\sum\limits_{j'=0}^{N-1}\left[\chi_{\Delta\Delta}(j,j')\Delta_{j'}^{(1)}+\chi_{\Delta M}(j,j')h_{j'}^{(1)}\right]\right\}, \\
h_j^{(1)}&=I_{\rm ff}\left\{(e\phi_0)\,\chi_{M\phi}(j,0)+\lambda\,\chi_{M\lambda}(j,0) \right.\\&\left.+\sum\limits_{j'=0}^{N-1}\left[\chi_{M\Delta}(j,j')\Delta_{j'}^{(1)}+\chi_{M M}(j,j')h_{j'}^{(1)}\right]\right\}.
\end{aligned}
\en
In these expressions we introduced the corresponding susceptibility functions
\beg\label{Susceptibilities}
\begin{aligned}
&\chi_{\alpha\beta}(j,j')=-T\sum\limits_{i\omega_n}\int\frac{d^2\bk_\parallel}{(2\pi)^2}\\&\times
{\rm Tr}\left[\hat{\cal O}_\alpha(j)\hat{\cal G}_{jj'}(\bk_\parallel,i\omega_n)\hat{\cal O}_\beta(j')\hat{\cal G}_{j'j}(\bk_\parallel,i\omega_n)\right],
\end{aligned}
\en
where $\hat{\cal G}_{jj'}(\bk_\parallel,i\omega_n)$ is the matrix propagator in the absence of the junction electric field, $\hat{H}_{\rm ext}=0$. 
The equations \eqref{LinCorr} form a closed system of linear equations for the components of vector ${\vec \psi}_j^{(1)}$, and several comments are in order. We note that two of its properties follow directly from symmetry. First, the operators $\hat{\cal O}_\Delta$ and $\hat{\cal O}_M$ are odd under time reversal, whereas $\hat{\cal O}_\phi$ and $\hat{\cal O}_\lambda$ are even. Thus in the time-reversal symmetric state the functions $\chi_{\Delta\phi}$, $\chi_{\Delta\lambda}$, $\chi_{M\phi}$ and $\chi_{M\lambda}$ vanish identically, so that the susceptibilities \eqref{Susceptibilities} must be evaluated in the ordered state, with ${\vec \psi}_j^{(0)}\not=0$ inserted into Eq.~\eqref{Hmf}, and ${\vec \psi}_j^{(1)}$ is odd under the reversal of the magnetic domain. Second, as $T\to T_c$ the homogeneous part of Eqs.~\eqref{LinCorr} reduces to the linearized self-consistency equations of Ref.~\cite{Brian2026}, whose largest eigenvalue equals unity at the transition. It also follows that the matrix of the system \eqref{LinCorr} becomes singular at $T_c$ and ${\vec \psi}_j^{(1)}\propto1/\sqrt{T_c-T}$ in its vicinity, in agreement with the arguments of Section~II.

\subsection{Equilibrium spin polarization contribution vs. magneto-electric response}
\begin{figure}
\includegraphics[width=\linewidth]{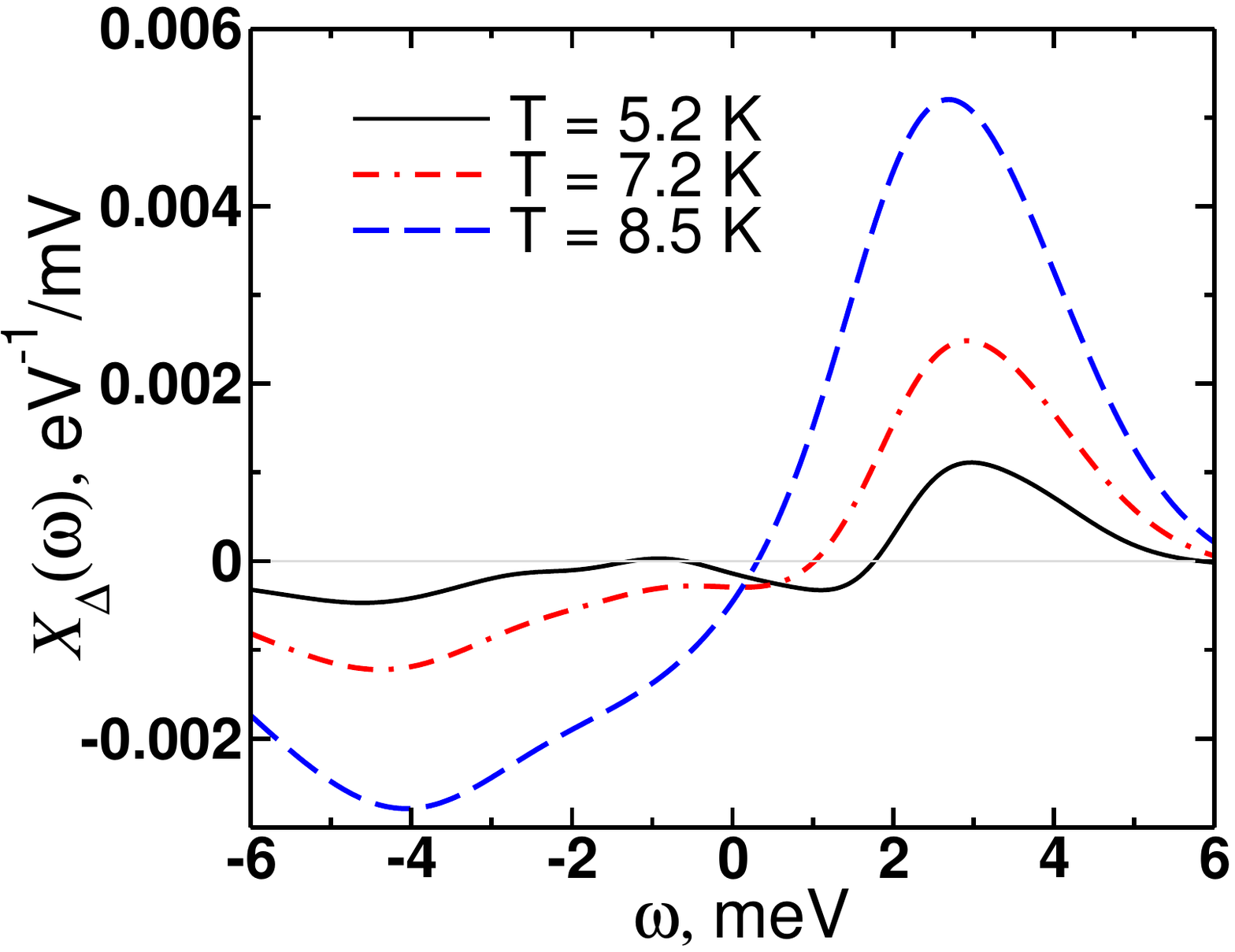}
\caption{Field-induced change of the spin-resolved spectral function of the apex due to the order-parameter channel, $X_\Delta(\omega)$, Eq.~\eqref{BDelta}, convolved with the thermal factor $-\partial f/\partial\omega$ at three different temperatures $T=5.2$, $7.2$ and $8.5$~K with 
$T_c\approx 9.8$~K. The results are shown per millivolt of bias for the unscreened field ($\kappa=1$) at the tip--substrate distance $d=7$~\AA, with $\ell_{\rm s}=0.021$~\AA\ and ${\mathsf z}_{\rm df}=0.1$~\AA. Per millivolt of bias $X_\Delta$ is smaller than the equilibrium spectrum $\rho_{\rm s}$ by three to four orders of magnitude.}
\label{Fig-BDelta}
\end{figure}

After the functions \eqref{Susceptibilities} have been computed, it becomes straightforward to solve the linear system \eqref{LinCorr} which, in turn, immediately allows us to evaluate $d{\vec \psi}_j^{(1)}/d\lambda$, $d{\vec \psi}_j^{(1)}/d(e\phi_0)$ and then function $X_\Delta(\omega)$.
In Fig. \ref{Fig-BDelta} we show the result of our calculation of the frequency dependence of the function $X_\Delta(\omega)$ at different temperatures. 
Specifically, $X_\Delta(\omega)$ increases as the temperature approaches $T_c$, reflecting the growth of the response of the order parameter, Eqs.~\eqref{LinCorr}, but even at $T=8.5$~K it remains three orders of magnitude smaller than $\rho_{\rm s}(\omega)$ per millivolt of bias. Furthermore, as $T\to T_c$ the energy dependence of $X_\Delta(\omega)$ approaches that of the equilibrium spectrum, and at $T=8.5$~K we find $X_\Delta(\omega)\approx\rho_{\rm s}(\omega)\,d\ln\Delta_{j=0}/dV$ with $d\ln\Delta_{j=0}/dV\approx10^{-3}$~mV$^{-1}$. Thus near $T_c$ the electric field changes the amplitude of the order parameter by the same fraction on every layer, leaving its profile and the ratio $h/\Delta$ unchanged, and since $\rho_{\rm s}$ is proportional to this amplitude, $X_\Delta(\omega)$ is $\rho_{\rm s}(\omega)$ multiplied by the relative change of the amplitude per unit bias.

\begin{figure}
\includegraphics[width=\linewidth]{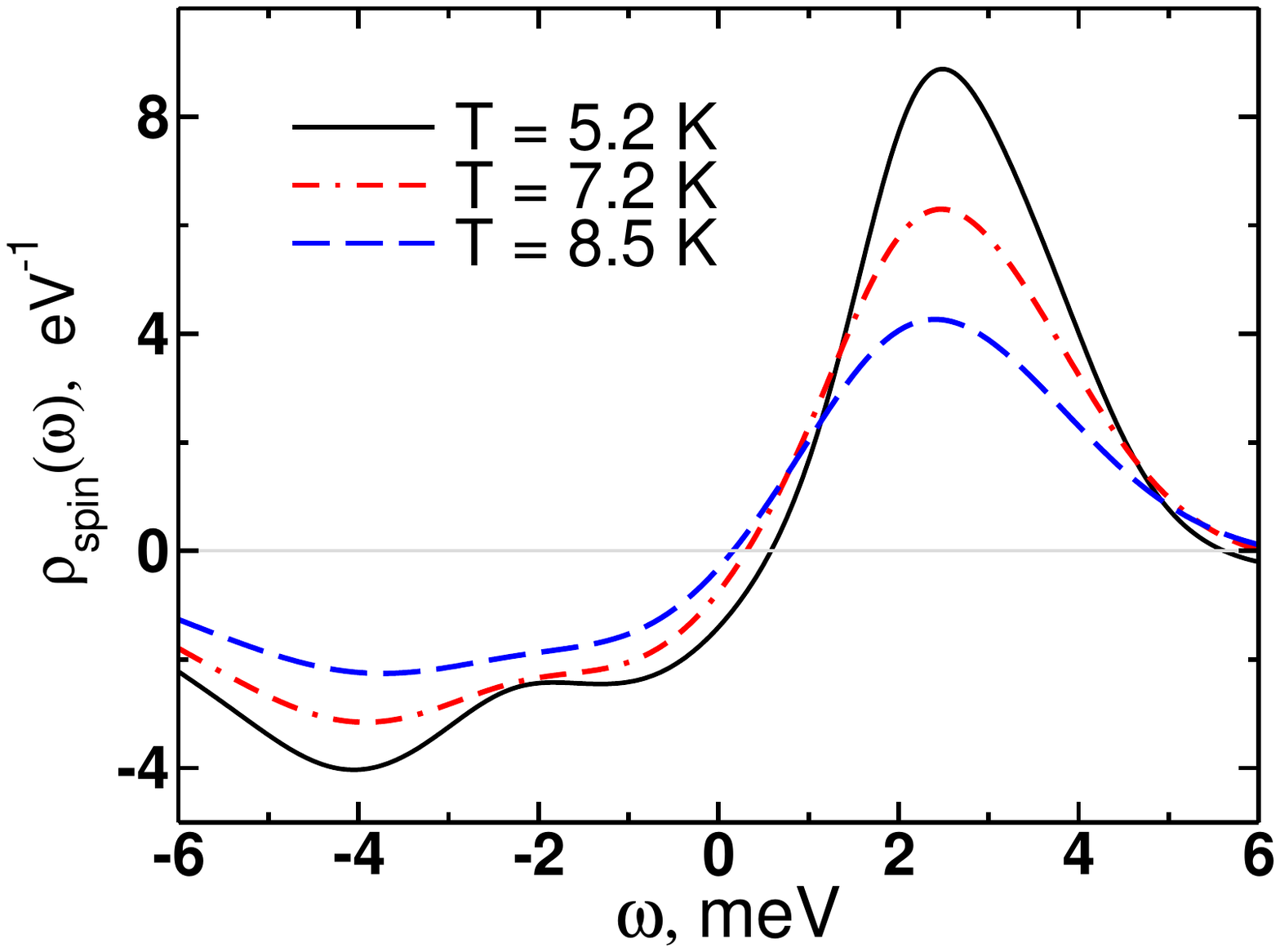}
\caption{Magnetic contrast part of the differential tunneling conductance, $\rho_{\rm spin}(V)=g_{\rm spin}(V)/e$, Eq.~\eqref{gspin}, as a function of $\omega=eV$ at $T=5.2$, $7.2$ and $8.5$~K. The field-induced contribution is that of the order-parameter channel $X_\Delta$, Fig.~\ref{Fig-BDelta}, for the unscreened field at $d=7$~\AA\ and $V_c=0$. On the scale of the figure, the curves coincide with the equilibrium contribution $g_{\rm s}(V)/e$: the field-induced correction does not exceed $1\%$ of the equilibrium part at any of the three temperatures.}
\label{Fig_rhospin}
\end{figure}

In Fig. \ref{Fig_rhospin} we show the frequency dependence of the magnetic contrast part of the spectral function $\rho_{\rm spin}(\omega)$, Eq.~\eqref{gspin}, which includes the field-induced contribution of the order-parameter channel $X_\Delta$ (Fig.~\ref{Fig-BDelta}), at three temperatures and for $V_c=0$. On the scale of the figure, the curves cannot be distinguished from the equilibrium contribution $g_{\rm s}(V)$: the field-induced correction does not exceed $1\%$ of the equilibrium part even at $T=8.5$~K, where $X_\Delta$ is the largest. Furthermore, the sign change of $g_{\rm spin}(V)$ at $eV^*$ moves from $0.58$~meV at $T=5.2$~K to $0.16$~meV at $T=8.5$~K, while the electric field shifts it by less than $1~\mu$eV at each temperature. 

\section{Discussion}\label{sec:discussion}
It is instructive to understand why the field-induced contributions turn out to be so small. The smallness of the dipole and gate channels, $X_d+X_{\rm g}$, has been explained in Section~II: of each millivolt of bias only about $10~\mu$eV reaches the apex, since the bias drops across the vacuum gap $d\approx7$~\AA\ while the field does work on the apex ion only over ${\mathsf z}_{\rm df}\approx0.1$~\AA\ and $\ell_{\rm s}\approx0.02$~\AA, and this energy is small compared with the width of the spectral features of the apex, which is of the order of $1$~meV. 

For the order-parameter channel $X_\Delta$, the suppression is of a different origin. As we have noted in Section \ref{Model}, the susceptibilities $\chi_{\Delta\phi}$, $\chi_{\Delta\lambda}$, $\chi_{M\phi}$ and $\chi_{M\lambda}$ (and therefore the field-induced corrections to $\Delta_{PT}$ and $h$) are odd functions of the order parameters: they vanish in the time-reversal symmetric state and change sign when the domain is reversed. Furthermore, at Matsubara frequencies large compared to $\Delta_{j=0}$ the order parameter acts as a small perturbation, so that the corresponding contribution to $\chi_{\Delta\phi}$ is the one to $\chi_{\Delta\Delta}$ suppressed by a factor of order $\Delta_{j=0}/\omega_n$. Thus $\chi_{\Delta\phi}$ is built from the states within about $10$~meV of the Fermi level, whose spectrum is reconstructed by the ordered state, while $\chi_{\Delta\Delta}$ collects contributions from the whole hybridized band. This is why the bare coupling $U_{\rm df}\chi_{\Delta\phi}(0,0)$ is more than an order of magnitude smaller than $U_{\rm df}\chi_{\Delta\Delta}(0,0)$, which is of order unity. The feedback of the ordered state, Eq.~\eqref{LinCorr}, enhances this coupling only by an order of magnitude, so that within the window $|eV|\lesssim 10$~meV the magnitude of $X_\Delta$ is only a few percent of that of $X_d+X_{\rm g}$. Thus, in order for the $X_\Delta$ channel to compete with the equilibrium spectrum within the experimental window, the enhancement would have to reach $10^3$--$10^4$, which in mean-field theory confines such a regime to a near vicinity of $T_c$,  when $T_c-T\lesssim10^{-3}T_c$.

Let us now comment on the electrostatics of the junction, which enters our results only through the screening factor $\kappa$ and the two length scales $\ell_{\rm s}$ and ${\mathsf z}_{\rm df}$, Eqs.~(\ref{Bdirect},\ref{dpsidV}). At this point we have not done a self-consistent Poisson--Hartree treatment of the charge induced at the apex. The reason is that our estimates show that it would not change our conclusions in any significant way. Specifically, the screening length $\ell_{\rm s}=\varepsilon_0a^2/(e^2\nu_{F0})\approx0.02$~\AA, where $\nu_{F0}\approx4.4$~eV$^{-1}$ is the density of states of the apex layer at the Fermi level, is itself the Hartree result in the limit in which the induced charge resides on the apex layer only. {Note that $\ell_{\rm s}$ is defined here as the ratio $\varepsilon_0a^2/e^2\nu_{F0}$ for a single layer treated as a charged sheet, not as a continuum Thomas--Fermi screening length; it only enters the gating term, which is subdominant.} Allowing the charge to spread over the subsequent layers corrects it by a factor $1+O(\varepsilon_0a/e^2D)$, i.e.\ by about $0.5\%$, so that for the gating channel $\kappa\approx1$. The field which acts on the $d$--$f$ dipole is the gradient of the potential between the vacuum and the first screened layer, which is about one half of the field in the vacuum gap, so that for the dipole channel $\kappa\approx2$. Since ${\mathsf z}_{\rm df}/\ell_{\rm s}\approx5$, the dipole term in Eq.~\eqref{dpsidV} dominates the drive of $X_\Delta$ unless the susceptibilities $\chi_{\Delta\phi}$ and $\chi_{M\phi}$ exceed $\chi_{\Delta\lambda}$ and $\chi_{M\lambda}$ by the same factor. Lastly, we note that the contact potential $V_c\sim0.1$--$1$~V produces a field of $10^{8}$--$10^{9}$~V/m in the gap at zero bias, for which $e\phi_0\approx0.3$--$3$~meV and $\lambda\approx0.7$--$7$~meV are comparable to $U_0$ and $\Delta_0$. {For an SmB$_6$ tip, whose work function $\Phi\approx3$~eV is unusually low, facing Fe$_{1+x}$Te with $\Phi\approx4.5$--$5$~eV, one expects $V_c$ at the upper end of this range or above, $V_c\approx1$--$2$~V, so that $\lambda\approx7$--$14$~meV and $e\phi_0\approx3$--$6$~meV at $d=7$~\AA, i.e.\ of the order of $\Delta_0$ or larger.} Thus the equilibrium state of the junction at $V=0$ is the ordered state in the contact field rather than the state at zero field, and Eq.~\eqref{eq:split} must be understood as the expansion about the {former}. {Its calculation requires a self-consistent treatment of the order parameter in a static field of order $10^{9}$~V/m, which we leave for future work. We stress that this does not affect the bias-odd part of the contrast, which is what reverses with the polarity and which remains proportional to $V/V_c$.}

Our results suggest a direct experimental test of which of the two proposed mechanisms produces the bias reversal. The field-induced part of the spin-resolved spectral function is proportional to the field at the apex and therefore to $1/d$, whereas the equilibrium part is not. In our calculation the field of the bias shifts $V^*$ by only a few $\mu$eV, so that $V^*$ should not change measurably when the set-point current is varied and with it the distance $d$. A field-induced moment large enough to determine the magnetic contrast, on the other hand, would make $V^*$ move with the set-point. Since the contact field also scales as $1/d$, a set-point dependence of $V^*$ would show that the electric field of the junction plays a role, while its absence would show that the bias reversal is a property of the equilibrium surface order. Thus measuring $V^*$ at several set-point currents is the most direct way to distinguish between the two mechanisms.

Finally, we would like to briefly comment on the contact potential $V_c$, which we have set to zero from the start. For $V_c\sim0.1$--$1$~V the contact potential exceeds the biases at which the magnetic contrast has been measured, $|V|\lesssim50$~mV \cite{Aishwarya2022,Banerjee2025}, so that the electric field in the gap, $E=(V+V_c)/d$, does not change its sign when the polarity of the bias is reversed. It follows that a magnetic moment which is linear in the field cannot change its sign with the polarity of the bias: only its bias-dependent part does, and this part is smaller than the one induced by the contact field by the factor $|V/V_c|$. Thus, for such values of $V_c$, the reversal of the magnetic contrast cannot be attributed to the reversal of a field-induced moment. In contrast, the reversal produced by the equilibrium spectrum does not rely on the sign of the field and, consequently, on the value of $V_c$ because it originates from the sign change of $\rho_{\rm s}(eV)$ itself. The contact field may, however, modify the equilibrium state itself and thereby shift the position of the critical voltage $V^*$ where the magnetic contrast part of the tunneling conductance changes sign.

\section{Conclusions}
In conclusion we would like to emphasize that, independently of whether the microscopic mechanism of time-reversal symmetry breaking (TRSB) proposed in Ref.~\cite{Brian2026} is realized, the observation of magnetic contrast \cite{Aishwarya2022,Banerjee2025} implies that time-reversal symmetry is broken at the surface of the SmB$_6$ tip \cite{Banerjee2025,Brian2026}. The point group of the (001) surface allows a time-reversal-odd order with an out-of-plane component, which acts as a mass for the surface Dirac cones and therefore gaps them out. With such an order the magnetic tunneling contrast emerges below the critical temperature of the TRSB transition, as we have illustrated in Fig.~\ref{Fig-polarity}. Consequently, the magnetic contrast part of the differential tunneling conductance is given by the sum of two terms: the first one, $g_{\rm s}(V)$, is determined by the equilibrium spin-resolved spectrum $\rho_{\rm s}(\omega)$ of the ordered apex, while the second one originates from the change of this spectrum by the electric field of the junction.  We would also like to stress that a bias spectrum taken at a fixed tip--substrate distance cannot separate these two terms. Furthermore, the ratio of the zero-bias intercept to the response extracted by the inversion procedure of Ref.~\cite{Banerjee2025} is equal to $-2V^*$, independently of how the spin-resolved spectral function is split between the equilibrium and the field-induced parts. Thus this ratio measures the position of the sign change of the magnetic contrast rather than the strength of the magneto-electric response. 

We note that our conclusions do not require the Fermi level to lie inside the gap of a surface Dirac cone. In the ordered state of Ref.~\cite{Brian2026} the Fermi level lies inside the gap of the $\bar\Gamma$ cone but above the gaps of the $\bar X$ and $\bar Y$ cones, so that the surface remains metallic. The sign changes of $\rho_{\rm s}(\omega)$ are set by the competition between the cones with masses of opposite signs, not by the positions of their gaps. Furthermore, the smallness of the field-induced contribution follows from ${\mathsf z}_{\rm df},\ell_{\rm s}\ll d$ and from the fact that the cross susceptibilities are odd in the order parameter, neither of which requires the Fermi level to lie inside a gap. 
 
Our microscopic calculation shows that within the experimental window $|eV|\lesssim10$~meV the field-induced term is much smaller than the equilibrium one. Specifically, within the framework of our mean-field model \cite{Brian2026} we find that the direct coupling of the electric field to the apex through the $d$--$f$ dipole and the electrostatic gating changes $\rho_{\rm s}(\omega)$ by about one percent per millivolt at the most, while the field-induced change of the order parameter yields a correction which does not exceed $1\%$ of the equilibrium contribution at the three experimentally relevant range of temperatures $T\sim T_c$ we have studied. It thus follows that both the energy dependence of the magnetic contrast and its reversal with bias are primarily determined by the equilibrium spin-resolved spectrum of the apex. It is for this reason that any assumption which results in ignoring the energy dependence of $g_{\rm s}(V)$ cannot be justified. Finally, we also find that the order-parameter channel grows as the temperature approaches $T_c$, and it can compete with the equilibrium contribution only in a narrow vicinity of the transition, however in that case the induced magnetic moment becomes negligibly  small. 

\begin{acknowledgments}
The work of MD was financially supported by the National Science
Foundation Grant No. DMR-2400484. 
The authors acknowledge the use of Anthropic's
Claude for cross-checking the analytical derivations, for independent
numerical verification of the results, and for figure preparation and
manuscript checking. All outputs of Claude were verified and validated by
the authors.
\end{acknowledgments}

\section*{Data availability} The data that support the findings of this article were generated by numerical evaluation of the analytical expressions derived herein and can be regenerated in full from the information given in the paper. For this reason the authors did not  deposit them in a public repository. The data and the codes that produce figures presented in this manuscript are available from the authors upon request. 

\begin{appendix}
\section{The $\Gamma_8$ quartets and the projector $\hat{\cal P}_2$}\label{AppendixG8}
Throughout this paper operators acting on the eight-component spinor \eqref{basis} are written as
$\ttau{i}\otimes\hs_j\otimes\hat\gamma_k$ ($i,j,k=0,\dots,3$), where the Pauli matrices $\ttau{i}$ act on the $(d,f)$ index,
$\hs_j$ act on the Kramers index $\sigma=\up,\dn$ and $\hat\gamma_k$ act on the orbital index $\alpha=1,2$ which labels the
two Kramers doublets of each quartet. The projectors on the two doublets are
\beg\label{P12}
\hat{\cal P}_{1}=\tfrac12\left(\hat\gamma_0+\hat\gamma_3\right),\qquad \hat{\cal P}_{2}=\tfrac12\left(\hat\gamma_0-\hat\gamma_3\right).
\en
In the basis \eqref{basis} the $8\times8$ matrix of $\ttau{i}\otimes\hs_j\otimes\hat\gamma_k$ is the Kronecker product $\ttau{i}\otimes\hat\gamma_k\otimes\hs_j$.
In this notation time reversal acts as $\Theta=\big(\ttau{0}\otimes i\hs_y\otimes\hat\gamma_0\big)K$, with $K$ the complex
conjugation.

In the cubic crystal field the $J=5/2$ multiplet of the $4f$ shell splits into a $\Gamma_7$
doublet and a $\Gamma_8$ quartet, and the latter consists of the two Kramers doublets
\cite{Alexandrov2013,Roy2014}
\beg\label{G8f}
\begin{aligned}
|\Gamma_8^{(1)},\pm\rangle&=\sqrt{\tfrac56}\,\big|\tfrac52,\pm\tfrac52\big\rangle+\sqrt{\tfrac16}\,\big|\tfrac52,\mp\tfrac32\big\rangle,\\
|\Gamma_8^{(2)},\pm\rangle&=\big|\tfrac52,\pm\tfrac12\big\rangle,
\end{aligned}
\en
where $|J,J_z\rangle$ are the states of total angular momentum $J=5/2$ formed from the orbital momentum $l=3$ and spin $1/2$.
In terms of the orbital and spin projections, $|l,m\rangle|\sigma\rangle$, the second doublet reads
\beg\label{G8f2}
|\Gamma_8^{(2)},\pm\rangle=\mp\sqrt{\tfrac37}\,|3,0\rangle|\!\pm\rangle\pm\sqrt{\tfrac47}\,|3,\pm1\rangle|\!\mp\rangle,
\en
while $|\Gamma_8^{(1)},\pm\rangle$ contains only the components with $m=\pm2$, $m=\pm3$ and $m=\mp1$. Since $Y_{lm}(\theta,\phi)$ vanishes on the
$z$-axis for every $m\neq0$, the charge density of $\Gamma_8^{(1)}$ vanishes along the $z$-axis whereas $\Gamma_8^{(2)}$, owing
to its $m=0$ component of weight $3/7$, extends along it. The $5d$ states form the $e_g$ quartet $e_g\otimes\tfrac12$,
\beg\label{G8d}
\begin{aligned}
|d_{1\sigma}\rangle&=\frac{|2,2\rangle+|2,-2\rangle}{\sqrt2}\,|\sigma\rangle\equiv d_{x^2-y^2}|\sigma\rangle,\\
|d_{2\sigma}\rangle&=|2,0\rangle|\sigma\rangle\equiv d_{3z^2-r^2}|\sigma\rangle,
\end{aligned}
\en
which transforms as $\Gamma_8$ as well. Rotations about the $z$-axis by $\pi/2$ conserve $J_z$ modulo $4$. Therefore
$d_{3z^2-r^2}|\sigma\rangle$, with $J_z=\pm1/2$, can mix with $\Gamma_8^{(2)}$ along the $z$-axis, while $d_{x^2-y^2}|\sigma\rangle$, with
$J_z\in\{5/2,-3/2\}$ for $\sigma=\up$ and $J_z\in\{-5/2,3/2\}$ for $\sigma=\dn$, can mix only with $\Gamma_8^{(1)}$. This fixes the
correspondence between the orbital index $\alpha$ of the $d$- and $f$-states in \eqref{basis}.

The states \eqref{G8f} are defined with the standard phases of the Clebsch--Gordan coefficients and of the spherical harmonics.
With these phases time reversal acts on the $d$-doublets as $i\hs_y K$ but on the $f$-doublets as $-i\hs_y K$, which is a
consequence of the odd orbital momentum $l=3$. In order to have the same operator $\Theta$ for both, we use the $f$-states
\beg\label{fphase}
|f_{\alpha\sigma}\rangle=i\,|\Gamma_8^{(\alpha)},\sigma\rangle .
\en
This is the convention of the model \cite{Alexandrov2013,Roy2014}: the hybridization along the $z$-axis, which is odd under
$z\to-z$, is $V(z)=iV\,{\rm diag}(0,\hs_z)$ in real space \cite{Alexandrov2013}, so that $V_h(\bk)$ is built from
$\hs_\alpha\sin k_\alpha$ with real coefficients, with $V\hs_z\sin k_z$ in its $(2,2)$ block. With the standard phases the same
block would carry an additional factor $i$.

The electric field along the surface normal couples to the intra-atomic coordinate
$\zeta=r\cos\theta$ of the apex ion. The operator $\zeta$ conserves both the orbital projection $m$ and the spin, hence $J_z$, and
it has nonzero matrix elements only between states of opposite parity, i.e.\ between the $5d$ and the $4f$ states. It follows from
Eqs.~\eqref{G8f}--\eqref{G8d} that $\zeta$ connects $d_{3z^2-r^2}|\sigma\rangle$ with $|\Gamma_8^{(2)},\sigma\rangle$ and
$d_{x^2-y^2}|\sigma\rangle$ with $|\Gamma_8^{(1)},\sigma\rangle$ only, and that it is diagonal in the Kramers index. Using
$\langle 2,m|\cos\theta|3,m\rangle=\sqrt{(9-m^2)/35}$ we find
\beg\label{zetaME}
\begin{aligned}
\langle d_{2\sigma}|\,\zeta\,|\Gamma_8^{(2)},\sigma\rangle&=-\sigma\,\frac{9}{35}\sqrt{\frac{5}{3}}\;{\cal R}_{5d,4f},\\
\langle d_{1\sigma}|\,\zeta\,|\Gamma_8^{(1)},\sigma\rangle&=-\sigma\,\frac{1}{7}\sqrt{\frac{5}{3}}\;{\cal R}_{5d,4f},
\end{aligned}
\en
where $\sigma=\pm1$ for $\up,\dn$ correspondingly and ${\cal R}_{5d,4f}=\int_0^\infty R_{5d}(r)R_{4f}(r)\,r^3dr$ is the radial integral. The opposite signs for the two Kramers partners, which follow from the opposite signs of the $|3,0\rangle$ component in Eq.~\eqref{G8f2}, give rise to the factor $\hs_z$. With the phase convention \eqref{fphase} the matrix elements \eqref{zetaME} acquire the factor
$i$, so that the $d$--$f$ part of $\zeta$ becomes
\beg\label{Dzfull}
\hat\zeta_{df}={\mathsf z}_{\rm df}\;\ttau{2}\otimes\hs_z\otimes\left(\hat{\cal P}_2+r\,\hat{\cal P}_1\right),
\en
with ${\mathsf z}_{\rm df}=(9/35)\sqrt{5/3}\,{\cal R}_{5d,4f}$ and $r=5/9$.
The factor $\ttau{2}$ instead of $\ttau{1}$ thus reflects the relative phase $i$ of the $d$- and $f$-states, which is the same
phase that appears in $V(z)$. Operator \eqref{Dzfull} is Hermitian, even under time reversal, odd under inversion and it is
invariant under the operations of $C_{4v}$, as required for the coupling to $E_z$.

In Eq.~\eqref{Hext} of the main text we have kept only the first term in Eq.~\eqref{Dzfull}, so that the
dipole acts within the doublet $\Gamma_8^{(2)}$ which extends along the surface normal. This is the same approximation which defines
the hybridization along the $z$-axis in the model \cite{Alexandrov2013,Roy2014}: there the overlap of the $\Gamma_8^{(1)}$ orbitals,
which extend perpendicular to the $z$-axis, is neglected and only $\Gamma_8^{(2)}$ hybridizes along $z$ \cite{Alexandrov2013}.
Indeed, the part of $H_{\rm bulk}(\bk)$ linear in $k_z$ is $Vk_z\,(\ttau{1}\otimes\hs_z\otimes\hat{\cal P}_2)$, and the dipole
operator $\ttau{2}\otimes\hs_z\otimes\hat{\cal P}_2$ connects the same states with the same spin structure. However, the term with $\hat{\cal P}_1$ is not small, since for the atomic states \eqref{G8f} and \eqref{G8d} its coefficient is $r=5/9$. Since
\beg\label{anticomm}
\left\{\hat{\cal O}_\Delta,\,\ttau{2}\otimes\hs_z\otimes\hat{\cal P}_\alpha\right\}=2\,\ttau{0}\otimes\hs_z\otimes\hat{\cal P}_\alpha,
\en
both terms of Eq.~\eqref{Dzfull} produce spin polarization in the ordered state in the same way, and the results for the full
operator \eqref{Dzfull} are obtained by adding the response to $\ttau{2}\otimes\hs_z\otimes\hat{\cal P}_1$ with the weight $r$.

\end{appendix}

\bibliography{references2}

\end{document}